\documentclass[conference]{IEEEtran}

\ifCLASSINFOpdf
\else
\fi
\usepackage[cmex10]{amsmath}

\usepackage{cite}
\usepackage{graphicx,epic,eepic,epsfig,amsmath,latexsym,amssymb,mathrsfs,verbatim,subfigure,color}

\newtheorem{definition}{Definition}

\newtheorem{lemma}[definition]{Lemma}

\newtheorem{theorem}[definition]{Theorem}
\newtheorem{corollary}[definition]{Corollary}

\newtheorem{example}[definition]{Example}

\newcommand{\bra}[1]{\langle#1|}
\newcommand{\ket}[1]{|#1\rangle}

\def\squareforqed{\hbox{\rlap{$\sqcap$}$\sqcup$}}
\def\qed{\ifmmode\squareforqed\else{\unskip\nobreak\hfil
\penalty50\hskip1em\null\nobreak\hfil\squareforqed
\parfillskip=0pt\finalhyphendemerits=0\endgraf}\fi}
\def\endenv{\ifmmode\;\else{\unskip\nobreak\hfil
\penalty50\hskip1em\null\nobreak\hfil\;
\parfillskip=0pt\finalhyphendemerits=0\endgraf}\fi}

\def\wt{\mathop{\rm wt}\nolimits}
\long\def\ignore#1{}

\begin{document}
%
\title{Concatenated Codes for Amplitude Damping}

\author{\IEEEauthorblockN{Tyler Jackson\IEEEauthorrefmark{1}\IEEEauthorrefmark{2}, Markus Grassl\IEEEauthorrefmark{3,4}, and Bei Zeng\IEEEauthorrefmark{1}\IEEEauthorrefmark{2}\IEEEauthorrefmark{5}}\medskip
\IEEEauthorblockA{
\IEEEauthorrefmark{1}Department of Mathematics $\&$ Statistics,
University of Guelph, Guelph, ON, N1G 2W1, Canada\\
\IEEEauthorrefmark{2}Institute for Quantum Computing, University of Waterloo,
Waterloo,  Ontario, N2L 3G1, Canada\\
\IEEEauthorrefmark{3}Institut f\"ur Optik, Information und Photonik,
  Universit\"at Erlangen-N\"urnberg, 91058 Erlangen, Germany\\
\IEEEauthorrefmark{4}Max Planck Institute for the Science of Light, Leuchs Division, 91058 Erlangen,
  Germany\\
\IEEEauthorrefmark{5} Canadian Institute for Advanced Research, Toronto, 
  Ontario, M5G 1Z8, Canada\\}
}


%


\maketitle


\maketitle

\begin{abstract}
We discuss a method to construct quantum 
codes correcting amplitude damping errors via code
concatenation. The inner codes
are chosen as asymmetric Calderbank-Shor-Steane (CSS) codes.
By concatenating with outer codes correcting symmetric errors,
many new codes with good parameters are found,
which are better than the amplitude damping codes 
obtained by any previously known construction.
\end{abstract}


\section{Introduction}
\label{sec:intro}
Channels transmitting quantum information represented by the density
matrix $\rho$ are completely positive, trace-preserving linear maps.
They can be represented in the Kraus decomposition
$\mathcal{A}(\rho)=\sum_k A_k\rho A_k^{\dag} $ with $\sum_k
A_k^{\dag}A_k=I$~\cite{nielsenchuang}.  The matrices $A_i$ are called
the Kraus operators or error set of the channel $\mathcal{A}$.

Most quantum error-correcting codes constructed so far are for the
depolarizing channel
\begin{equation}
\mathcal{A}_{\text{DP}}(\rho)=(1-p)\rho +\frac{p}{3}(X\rho X+Y\rho Y +Z\rho Z),
\end{equation}
where the Pauli $X,Y,Z$ errors happen equally likely. The Kraus
operators of $\mathcal{A}_{\text{DP}}$ are $\{\sqrt{1-p}I,
\sqrt{\frac{p}{3}}X, \sqrt{\frac{p}{3}}Y, \sqrt{\frac{p}{3}}Z\}$.

The assumption of equal probability for the Pauli $X,Y,Z$ errors in
fact models the worst case scenario of `white noise', where all kind
of errors happen.  However, in practical systems, some errors are
usually more likely to happen than others.  A more realistic error
model for physical systems is based on the common noise processes
described by amplitude damping and phase damping.  The corresponding
asymmetric channel is given by
\begin{alignat}{5}
\label{eq:AS}
\mathcal{A}_{\text{AS}}(\rho)={}&(1-(2p_{xy}+p_z))\rho\nonumber\\
&{}+p_{xy}(X\rho X+Y\rho Y)+p_z Z\rho Z,
\end{alignat}
where the Pauli $X$ and $Y$ errors happen with equal probability
$p_{xy}$, which is determined by the amplitude damping (AD) noise.
The probability $p_z$ of the Pauli $Z$ error depends on the phase
damping noise, and in general we have $p_{xy}\neq p_z$.

The amplitude damping channel is given by~\cite{nielsenchuang}
\begin{equation}\label{eq:ADchannel}
\mathcal{A}_{\text{AD}}(\rho)=A_0\rho A_0^{\dag}+A_1\rho A_1^{\dag}, 
\end{equation}
where the Kraus operators are
\begin{equation}
\label{eq:AD}
A_0=\begin{pmatrix} 1 & 0 \\ 0 & \sqrt{1-\gamma} \end{pmatrix},
\quad
A_1=\begin{pmatrix} 0 & \sqrt{\gamma} \\ 0 & 0 \end{pmatrix},
\end{equation}
where $\gamma$ is a damping parameter.
The AD channel models, e.g., photon loss in optical fibers, or
spontaneous emission of atoms~\cite{nielsenchuang,chuang1997bosonic}.

It has been first demonstrated in~\cite{leung1997approximate} that
designing QECCs adaptively to the AD noise can result in better
codes. In particular, a four-qubit code correcting a single AD error
was found, using less qubits than the smallest single-error-correcting
code for the depolarizing channel that needs five
qubits~\cite{bennett1996mixed,laflamme1996perfect}. Generalizations of
this four-qubit code are discussed
in~\cite{thesis:gottesman,fletcher2008channel}. In particular, it was
discussed in ~\cite{thesis:gottesman} that Shor's nine-qubit code can
correct $2$ AD errors, despite the fact that the code only corrects a
single error for the depolarizing channel.

Subsequent works borrow ideas from the construction of classical
asymmetric codes~\cite{klove1981error}, combined with the codeword
stabilized (CWS) quantum code method~\cite{CSS+09}, to construct
single-error-correcting AD codes, including both stabilizer codes and
non-additive codes~\cite{lang2007nonadditive,shor2011high}.
Multi-error-correcting AD codes are discussed in~\cite{duan2010multi},
based on a concatenation method.  In particular, the inner code is
chosen as the two-qubit code $\{\ket{01},\ket{10}\}$ based on the
classical dual-rail code, which results in a quantum erasure channel
for the outer codes. Many good stabilizer AD codes are constructed by
concatenating with the quantum erasure codes. However, due to the
choice of the inner code, the rate of the constructed AD codes can
never exceed $1/2$.

In this work, we discuss a new method to construct AD codes via
concatenation.  We choose the inner codes as codes correcting certain
kind of asymmetric errors.  By carefully analyzing the error model for
the AD channel, we introduce the concept of `effective weight' for
errors and `effective distance' for the AD codes.  This allows us to
use outer codes correcting symmetric errors (i.e., the `usual' codes
designed to correct depolarizing errors).  Our new method results in
many new AD codes with good parameters, which are better than the AD
codes obtained by any previously known construction.

\section{Background}

A QECC $Q$ is a subspace of the space of $n$ qudits
$(\mathbb{C}^q)^{\otimes n}$, with single qudit dimension $q$.  For a
$K$-dimensional code space spanned by the orthonormal basis
$\{\ket{\psi_i}, i=1,\ldots,K\}$ and an error set $\mathcal{A}$, there
is a physical operation detecting all the elements
$A_{\mu}\in\mathcal{A}$ if the error detection
condition~\cite{bennett1996mixed,knill1997theory}
\begin{equation}
\label{eq:EDC}
\bra{\psi_i}A_{\mu}\ket{\psi_j}=c_{\mu}\delta_{ij}
\end{equation} 
is satisfied. 

The notation $((n,K))_q$ is used to denote a qudit QECC with length
$n$ and dimension $K$.  A stabilizer QECC has dimension $K=q^k$ for
some integer $k$, and the notation $[\![n,k]\!]_q$ is used to denote a
qudit stabilizer code with length $n$ and dimension $q^k$.  A code $Q$
is of distance $d$ if Eq.~\eqref{eq:EDC} is satisfied for all
$A_{\mu}$ that act nontrivially on at most $d-1$ qudits.

In this work, we focus on the construction of AD codes, which are qubit
codes with $q=2$. However, qudit codes with $q=2^r$ are used as outer
codes for the concatenation constructions to get qubit AD codes.

We consider error sets $\mathcal{A}$ of Pauli type.  For
multi-qubit Pauli operators, for instance, $X\otimes Y\otimes I\otimes
Z$, we will write it as $XYIZ$ or $X_1Y_2Z_4$ (where the subscripts
denote the number of the qubits that the Pauli $X,Y,Z$ operator is
acting on), when no confusion arises. For the AD channel
$\mathcal{A}_{\text{AD}}$ as given in Eq.~\eqref{eq:AD}, the Kraus
operators $A_0$ and $A_1$ are not Pauli operators. However, we can
find Pauli error models that lead to codes correcting AD errors.

Notice that
\begin{equation}
\label{eq:A1}
A_1=\frac{\sqrt{\gamma}}{2}(X+iY),\quad
A_0=I-\frac{\gamma}{4}(I-Z)+O(\gamma^2),
\end{equation}
and hence, $A_0$ which is of different order in $\gamma$ as $A_1$. So
the corresponding asymmetric error model as given in Eq.~\eqref{eq:AS}
has $p_{xy}\propto \gamma$ and $p_z\propto \gamma^2$.

A $t$-error-correcting AD code (or $t$-code in short) improves the
fidelity of the transmitted state from $1-\gamma$ to $1-\gamma^t$.
For instance, for $t=1$, we only need to correct a single $A_1$ error
and detect a single $A_0$ error~\cite{GWYZ14}.  In terms of Pauli operators, we
only need to correct a single $X$ and $Y$ error, and detect a single
$Z$ error. In other words, a code that detects the error set
$\mathcal{A}^{\{1\}}$ that is given by
\begin{equation}
\label{eq:EA1}
\mathcal{A}^{\{1\}}=\{I\}\cup\{X_i,Y_i,Z_i,\;X_iX_j,X_iY_j,Y_iY_j\},
\end{equation}
with $i,j \in [1,n]$,
is a $1$-code that corrects a single AD error.

Pauli error models that lead to codes correcting $t$ AD errors can be
given similarly.  For instance, codes detecting the Pauli
error set given by
$
\mathcal{A}^{\{2\}}=\{A_{\mu}A_{\nu}\colon A_{\mu},A_{\nu}\in\mathcal{A}^{\{1\}}\}
$
are $2$-codes that correct $2$ AD errors.
We will similarly denote by $\mathcal{A}^{\{t\}}$ the Pauli error set
that results in $t$-codes.

\section{Concatenated method}

We examine the weight properties of the elements in
$\mathcal{A}^{\{t\}}$, which leads to new effective error models that
are more convenient for constructing codes detecting the error set
$\mathcal{A}^{\{t\}}$. From Eq.~\eqref{eq:ADchannel} and
Eq.~\eqref{eq:A1} it follows that each $Z$ error contributes a factor
of $\gamma^2$ to the noise, while $X$ or $Y$ errors contribute a factor
of $\gamma$. In other words, when we consider each $X$ or $Y$
error as `$1$ error', then each $Z$ error will be `effectively $2$
errors'. Motivated by this observation, we have the following
definition for `effective weight'.

\begin{definition}
For any tensor product $E$ of Pauli errors, each tensor factor $X$ or
$Y$ has effective weight $1$, and each factor $Z$ has effective weight
$2$.  The effective weight of $E$ is the sum of the effective weight
of all factors $X,Y,Z$ in $E$, and is denoted by $\wt_e(E)$.
\end{definition}

As an example, for $E=XYIZ\in\mathcal{A}^{\{2\}}$, $\wt_e(E)=4$. In
fact, we have the following result on the effective weight of the
elements in $\mathcal{A}^{\{t\}}$.
\begin{lemma}
Any element $E\in\mathcal{A}^{\{t\}}$ has effective weight $\wt_e(E)\leq
2t$.
\end{lemma}
\begin{IEEEproof}
Notice that any element $E\in\mathcal{A}^{\{t\}}$ will be a product of
at most $t$ elements from $\mathcal{A}^{\{1\}}$ as given in
Eq.~\eqref{eq:EA1}.  Any element in $\mathcal{A}^{\{1\}}$ has at most
effective weight $2$, hence $E$ has at most effective weight $2t$.
\end{IEEEproof}

Obviously, the upper bound $2t$ is achievable by some elements
$E\in\mathcal{A}^{\{t\}}$.  We can now define the effective distance
$d_e$ for $t$-AD-error-correcting codes that detect the error set
$\mathcal{A}^{\{t\}}$. This effective distance will later allow us to
compare our new codes with codes for the depolarizing channel with the
usual code distance $d$ (i.e., each $X,Y,Z$ has weight $1$).

\begin{definition}
A code has effective distance $d_e=s$, if it detects Pauli errors
of effective weight up to $s-1$.
\end{definition}

Therefore, if a code has effective distance $d_e=2t+1$, then it detects
the error set $\mathcal{A}^{\{t\}}$, and is hence a $t$-code.

Now we are ready to present our concatenation method.

\begin{theorem}
\label{th:con}
Starting from an inner $[\![n_1,k_1]\!]_2$ code $\mathcal{Q}_{i}$ with
effective distance $d_e$, concatenation of an
$[\![n_2,k_2,\delta]\!]_{2^{k_1}}$ qudit outer code $\mathcal{Q}_{o}$
with distance $\delta$ results in a concatenated code
$[\![n_1n_2,k_1k_2]\!]_2$ with effective distance at least
$d_e\delta$.
\end{theorem}

\begin{IEEEproof}
The concatenated code $\mathcal{Q}$ is a stabilizer code with length
$n_1n_2$ and dimension $(2^{k_1})^{k_2}$, hence encoding $k_1k_2$
qubits. Denote the stabilizer of $\mathcal{Q}$ by
$S_{\mathcal{Q}}$. It has two sets of generators.  The first set is
obtained by replacing each tensor factor of the generators of the
stabilizer $S_{\mathcal{Q}_o}$ of the outer code by the corresponding
logical operator of the inner code. The second set is formed by the
stabilizer $S_{i}$ of the inner code acting on each block of $n_1$
qubits.

For the outer code $\mathcal{Q}_{o}$, any nontrivial logical operator
in $C(S_{\mathcal{Q}_{o}})\setminus S_{\mathcal{Q}_{o}}$ has weight at
least $\delta$, where $C(S)$ is the centralizer of the stabilizer
$S$. Likewise, the logical operators in
$C(S_{\mathcal{Q}_{i}})\setminus S_{\mathcal{Q}_{i}}$ of the inner
code have effective weight at least $d_e$.  The logical operators of
the concatenated code are obtained by replacing each tensor factor in
the logical operators of the outer code by the corresponding logical
operator of the inner code.  Those operators have effective weight at
least $d_e\delta$.  As for standard concatenation of quantum codes
\cite{KLZ98}, multiplying a logical operator of $\mathcal{Q}$ by an
element of the stabilizer $S_\mathcal{Q}$ will not result in an
effective weight less than $d_e\delta$.
\end{IEEEproof}

\section{The $[\![r,r-1]\!]_2$ inner code}

To examine the power of the construction for AD codes given in
Theorem~\ref{th:con}, we will start with simple inner codes.  We take
classical linear binary codes of distance $2$ with length $r$ and
dimension $r-1$ (hence cardinality $2^{r-1}$). For any length $r$,
such a distance-$2$ code will be formed by all bit strings of length
$r$ with even Hamming weight.  For any such classical code
$\mathcal{C}_r=[r,r-1,2]_2$, the corresponding quantum code
$\mathcal{Q}_r=[\![r,r-1]\!]_2$ is spanned by the computational basis
vectors $\ket{\mathbf{c}_i}$ for all
${\mathbf{c}_i}\in\mathcal{C}_r$. We first examine the effective
distance of $\mathcal{Q}_r$.

\begin{lemma}
The code $\mathcal{Q}_r$ defined above has effective distance $d_e=2$.
\end{lemma}

\begin{IEEEproof}
The only non-trivial element of the stabilizer $S_r$ of the code
$\mathcal{Q}_r$ is the $r$-fold tensor product $Z^{\otimes r}$.  We
need to look at the effective weights of the logical operators that
are in $C(S_r)\setminus S_r$, where $C(S_r)$ is the centralizer of
$S_r$.  These are Pauli operators that commute with $Z^{\otimes r}$.
Clearly, a single $Z$ (i.e., $Z_i$) operator having effective weight
two is in $C(S_r)\setminus S_r$, but this set does not contain a
single $X$ or $Y$ operator. The tensor product of two $X$ or $Y$
operators (i.e., $X_iX_j, X_iY_j, Y_iX_j, Y_iY_j$) is in
$C(S_r)\setminus S_r$. Therefore, every logical operator of
$\mathcal{Q}_r$ has effective weight at least two, and hence the
effective distance of $\mathcal{Q}_r$ is $2$.
\end{IEEEproof}

Since the dimension of the quantum code $\mathcal{Q}_r$ is $2^{r-1}$,
it can be used as inner code for the concatenation with a qudit outer
codes with single qudit dimension $q=2^{r-1}$. For the construction of
a $t$-code, we need effective distance $2t+1$ for the concatenated code.

\begin{theorem}
\label{th:main}
Given an $[\![n,k,\delta]\!]_{2^{r-1}}$ stabilizer code, a quantum code
$\mathcal{Q}$ with parameters $[\![rn-1,(r-1)k]\!]_2$ and effective
distance $d_e\ge 2\delta-1$ can be constructed. This is a $t$-code with
$t=\delta-1$.
\end{theorem}

\begin{IEEEproof}
We start from an $[\![n,k,\delta]\!]_{2^{r-1}}$ stabilizer code of
length $n$, and each qudit has dimension $2^{r-1}$. The first qudit is
encoded into a trivial qubit code with parameters
$[\![r-1,r-1,d_e=1]\!]_2$.  Each of the other qudits $j=2,3,\ldots,n$
is encoded into the code $\mathcal{Q}_r$ with parameters
$[\![r,r-1,d_e=2]\!]_2$. The resulting concatenated code $\mathcal{Q}$
is a stabilizer code of length $(r-1)+(n-1)r=rn-1$ and dimension
$(2^{r-1})^{k}$, hence encoding $(r-1)k$ qubits.  Any logical operator
of $[\![n,k,\delta]\!]_{2^{r-1}}$ has weight at least $\delta$.  Hence
any logical operator of $\mathcal{Q}$ that acts trivially on the first
qudit has effective weight at least $2\delta$.  Logical operators of
$\mathcal{Q}$ that act non-trivially on the first qudit have effective
weight at least $1+2(\delta-1)=2\delta-1$. Therefore, the effective
distance of $\mathcal{Q}$ is $d_e\ge 2\delta-1$.
\end{IEEEproof}

\begin{example}
\label{eg:9bit}
Starting from the $[\![5,1,3]\!]_2$ code with stabilizer generated by
\[\arraycolsep0.25\arraycolsep
\begin{array}{ccccc}
X & Z & Z & X & I \\
I & X & Z & Z & X \\
X & I & X & Z & Z \\
Z & X & I & X & Z
\end{array}
\]
and encoding qubits $2,3,4,5$ into the code $\mathcal{Q}_2$ stabilized
by $ZZ$, we get a $[\![9,1]\!]_2$ code with effective distance $d=2\cdot
3-1=5$, which corrects two AD errors. By choosing the logical
operators for $\mathcal{Q}_2$ as $\bar{X}=XX$ and $\bar{Z}=ZI$, the
stabilizer of the $[\![9,1]\!]_2$ code is generated by
\[\arraycolsep0.5\arraycolsep
\begin{array}{ccccc}
X & ZI & ZI & XX & II\\
I & XX & ZI & ZI & XX\\
X & II & XX & ZI & ZI\\
Z & XX & II & XX & ZI\\[0.5ex]
I & ZZ & II & II & II\\
I & II & ZZ & II & II\\
I & II & II & ZZ & II\\
I & II & II & II & ZZ
\end{array}
\]
\end{example}
Notice the two groups of generators as mentioned in the proof of Theorem~\ref{th:con}.

We remark that the $[\![9,1]\!]_2$ $2$-code given above is in fact
local Clifford equivalent to one of the $[\![9,1]\!]_2$ codes found
in~\cite{JGZ16} via exhaustive numerical search for CWS codes
detecting the error set $\mathcal{A}^{\{2\}}$. It is one of the best
$2$-codes known, which beats the $[\![10,1]\!]_2$ $2$-code found
in~\cite{duan2010multi}.  In fact, the construction
in~\cite{duan2010multi} can be viewed as a special case of
Theorem~\ref{th:main}, by concatenating all qudits of an outer code
with the inner code $\mathcal{Q}_2$.  Notice that
in~\cite{duan2010multi}, codes with effective distance $2\delta$ are
constructed in order to obtain $t$-codes with $t=\delta-1$, which
results in length $2n$ instead of $2n-1$ as given by
Theorem~\ref{th:main}.  In other words, by using
Theorem~\ref{th:main}, the length of any $t$-code constructed
in~\cite{duan2010multi} can be reduced by one.

For decoding, the inner code $[\![r,r-1]\!]_2$ will be used to detect
single $X$- and $Y$-errors. This provides side-information on detected
errors (erasures) for the outer code and allows to simultaneously
correct $e$ erasures and $f$ erroneous blocks with $r$ qubits each, as
long as $e+2f<\delta$.

\section{Parameters of New AD Codes}

In this section we discuss the parameters of the new AD codes found by
our concatenated method when using the inner code $\mathcal{Q}_r$.  We
compare the effective distance $d_e$ of the new codes constructed via
our concatenated method to the distance $d_{\text{lb}}$ of the best
known stabilizer codes.

The best possible parameters for our concatenation technique are
expected when the outer code is an optimal quantum code, and quantum
MDS (QMDS) codes in particular.  QMDS codes have parameters
$[\![n,n+2-2d,d]\!]_q$, i.e., they attain the quantum Singleton bound
$k+2d\le n+2$ \cite{knill1997theory,Rai99}.  QMDS codes are known to
exists for all $n\le q+1$, for $n=q^2-1,q^2,q^2+1$ and some $d\le
q+1$, as well as for many parameters $n\le q^2+1$, $d\le q+1$
\cite{GBR04}.  In general it seems as if for a qudit QMDS code with
qudit dimension $q$ we have the bounds $d\le q+1$, and $n\le q^2+1$,
with the exception of codes $[\![4^m+2,2^m-4,4]\!]_{2^m}$ (see
\cite{GrRo15}).

In order to construct a $t$-code, we use QMDS codes
$[\![n,n-2t,t+1]\!]_q$ where $q=2^{r-1}\ge t$ as outer code and the
code $\mathcal{Q}_r=[\![r,r-1]\!]_2$ as inner code, yielding a
$t$-code of length $rn-1$ encoding $(r-1)(n-2t)=rn-n-2rt+2t$
qubits.

The parameters of our codes based on the concatenation of QMDS codes
and the code $\mathcal{Q}_r$ are presented in Table~\ref{tab:QMDS}.
The last column labeled $d_{\text{lb}}$ lists the largest known lower
bound $d_{\text{lb}}$ on the minimum distance of a stabilizer code for
the depolarizing channel (see \cite{Grassl:codetables}). Here we
consider only codes of length up to $n_{\text{max}}=128$.  We only
list the parameters $[\![n,k,d_e=2t+1]\!]_2$ of $t$-codes for which
the effective distance $d_e$ exceeds the lower bound
$d_{\text{lb}}$ (i.e., $d_e>d_{\text{lb}}$).  Furthermore, we omit parameters for which we find
even betters codes (smaller length, larger dimension, or larger
effective distance).

In Tables~\ref{tab:nonQMDS2} and~\ref{tab:nonQMDS4_8} we list
parameters of the best $t$-codes we found using outer codes that do
not reach the quantum Singleton bound $k+2d\le n+2$, but have the
largest minimum distance among the known codes. The codes in
Table~\ref{tab:nonQMDS2} are based on qubit codes as outer codes and
hence comparable to the codes in~\cite{duan2010multi}, but 
reducing the length by one as discussed above.

\begin{table}[hbt]
\caption{Concatenated codes $[\![n,k,d_e]\!]_2$ for the AD channel
  based on QMDS outer codes with qudit dimension $2$, $4$, $8$, and $16$.}
\label{tab:QMDS}\def\arraystretch{1.17}
\centerline{$
\begin{array}{|c|c|c|c|}
\hline
t&\text{concatenated code} & \text{outer code} & d_{\text{lb}}\\
\hline
1 &[\![7,2,d_e=3]\!]_2 & [\![4,2,2]\!]_2 & 2 \\\hline
2 &[\![9,1,d_e=5]\!]_2 & [\![5,1,3]\!]_2 & 3 \\\hline
\multicolumn{4}{c@{}}{\rule{0pt}{10pt}}\\
\hline
t & \text{concatenated code} & \text{outer code} & d_{\text{lb}}\\
\hline
3 &[\![23,4,d_e=7]\!]_2 & [\![8,2,4]\!]_{2^2} & 6 \\{}
  &[\![26,6,d_e=7]\!]_2 & [\![9,3,4]\!]_{2^2} & 6 \\{}
  &[\![29,8,d_e=7]\!]_2 & [\![10,4,4]\!]_{2^2} & 6 \\{}
  &[\![41,16,d_e=7]\!]_2 & [\![14,8,4]\!]_{2^2} & 6 \\\hline
4 &[\![26,2,d_e=9]\!]_2 & [\![9,1,5]\!]_{2^2} & 8 \\{}
  &[\![50,18,d_e=9]\!]_2 & [\![17,9,5]\!]_{2^2} & 8 \\\hline
\multicolumn{4}{c@{}}{\rule{0pt}{10pt}}\\
\hline
t &\text{concatenated code} & \text{outer code} & d_{\text{lb}}\\
\hline
4 &[\![39,6,d_e=9]\!]_2 & [\![10,2,5]\!]_{2^3} & 8 \\{}
  &[\![43,9,d_e=9]\!]_2 & [\![11,3,5]\!]_{2^3} & 8 \\{}
  &[\![47,12,d_e=9]\!]_2 & [\![12,4,5]\!]_{2^3} & 8 \\{}
  &[\![59,21,d_e=9]\!]_2 & [\![15,7,5]\!]_{2^3} & 8 \\{}
  &[\![75,33,d_e=9]\!]_2 & [\![19,11,5]\!]_{2^3} & 8 \\\hline
5 &[\![47,6,d_e=11]\!]_2 & [\![12,2,6]\!]_{2^3} & 10 \\{}
  &[\![63,18,d_e=11]\!]_2 & [\![16,6,6]\!]_{2^3} & 10 \\{}
  &[\![71,24,d_e=11]\!]_2 & [\![18,8,6]\!]_{2^3} & 10 \\{}
  &[\![75,27,d_e=11]\!]_2 & [\![19,9,6]\!]_{2^3} & 10 \\{}
  &[\![79,30,d_e=11]\!]_2 & [\![20,10,6]\!]_{2^3} & 9 \\{}
  &[\![83,33,d_e=11]\!]_2 & [\![21,11,6]\!]_{2^3} & 10 \\{}
  &[\![91,39,d_e=11]\!]_2 & [\![23,13,6]\!]_{2^3} & 10 \\{}
  &[\![99,45,d_e=11]\!]_2 & [\![25,15,6]\!]_{2^3} & 10 \\{}
  &[\![103,48,d_e=11]\!]_2 & [\![26,16,6]\!]_{2^3} & 10 \\{}
  &[\![107,51,d_e=11]\!]_2 & [\![27,17,6]\!]_{2^3} & 10 \\{}
  &[\![111,54,d_e=11]\!]_2 & [\![28,18,6]\!]_{2^3} & 10 \\\hline
6 &[\![95,36,d_e=13]\!]_2 & [\![24,12,7]\!]_{2^3} & 12 \\{}
  &[\![99,39,d_e=13]\!]_2 & [\![25,13,7]\!]_{2^3} & 11 \\{}
  &[\![103,42,d_e=13]\!]_2 & [\![26,14,7]\!]_{2^3} & 11 \\{}
  &[\![107,45,d_e=13]\!]_2 & [\![27,15,7]\!]_{2^3} & 11 \\{}
  &[\![111,48,d_e=13]\!]_2 & [\![28,16,7]\!]_{2^3} & 11 \\{}
  &[\![115,51,d_e=13]\!]_2 & [\![29,17,7]\!]_{2^3} & 12 \\{}
  &[\![119,54,d_e=13]\!]_2 & [\![30,18,7]\!]_{2^3} & 12 \\{}
  &[\![123,57,d_e=13]\!]_2 & [\![31,19,7]\!]_{2^3} & 12 \\{}
  &[\![127,60,d_e=13]\!]_2 & [\![32,20,7]\!]_{2^3} & 11 \\\hline
7 &[\![127,54,d_e=15]\!]_2 & [\![32,18,8]\!]_{2^3} & 13 \\\hline
\multicolumn{4}{c@{}}{\rule{0pt}{10pt}}\\
\hline
t & \text{concatenated code} & \text{outer code} & d_{\text{lb}}\\
\hline
6 &[\![79,16,d_e=13]\!]_2 & [\![16,4,7]\!]_{2^4} & 12 \\\hline
7 &[\![119,40,d_e=15]\!]_2 & [\![24,10,8]\!]_{2^4} & 14\\\hline
\end{array}$}
\end{table}

\begin{table}[hbt]
\caption{Concatenated codes $[\![n,k,d_e]\!]_2$ for the AD channel
  based on non-QMDS outer qubit codes.}
\label{tab:nonQMDS2}\def\arraystretch{1.1}
\centerline{$
\begin{array}{|c|c|c|c|}
\hline
t &\text{concatenated code} & \text{outer code} & d_{\text{lb}}\\
\hline
3 &[\![19,2,d_e=7]\!]_2 & [\![10,2,4]\!]_2 & 6 \\{}
  &[\![23,4,d_e=7]\!]_2 & [\![12,4,4]\!]_2 & 6 \\\hline
4 &[\![21,1,d_e=9]\!]_2 & [\![11,1,5]\!]_2 & 7 \\{}
  &[\![31,4,d_e=9]\!]_2 & [\![16,4,5]\!]_2 & 8 \\{}
  &[\![35,6,d_e=9]\!]_2 & [\![18,6,5]\!]_2 & 8 \\\hline
5 &[\![31,2,d_e=11]\!]_2 & [\![16,2,6]\!]_2 & 10 \\{}
  &[\![39,4,d_e=11]\!]_2 & [\![20,4,6]\!]_2 & 9 \\{}
  &[\![41,5,d_e=11]\!]_2 & [\![21,5,6]\!]_2 & 9 \\{}
  &[\![47,6,d_e=11]\!]_2 & [\![24,6,6]\!]_2 & 10 \\{}
  &[\![55,12,d_e=11]\!]_2 & [\![28,12,6]\!]_2 & 10 \\\hline
6 &[\![33,1,d_e=13]\!]_2 & [\![17,1,7]\!]_2 & 11 \\{}
  &[\![47,3,d_e=13]\!]_2 & [\![24,3,7]\!]_2 & 11 \\{}
  &[\![49,5,d_e=13]\!]_2 & [\![25,5,7]\!]_2 & 11 \\{}
  &[\![59,8,d_e=13]\!]_2 & [\![30,8,7]\!]_2 & 12 \\{}
  &[\![63,10,d_e=13]\!]_2 & [\![32,10,7]\!]_2 & 12 \\\hline
7 &[\![47,1,d_e=15]\!]_2 & [\![24,1,8]\!]_2 & 13 \\{}
  &[\![51,4,d_e=15]\!]_2 & [\![26,4,8]\!]_2 & 12 \\{}
  &[\![59,5,d_e=15]\!]_2 & [\![30,5,8]\!]_2 & 13 \\{}
  &[\![63,6,d_e=15]\!]_2 & [\![32,6,8]\!]_2 & 14 \\{}
  &[\![65,7,d_e=15]\!]_2 & [\![33,7,8]\!]_2 & 13 \\{}
  &[\![67,8,d_e=15]\!]_2 & [\![34,8,8]\!]_2 & 14 \\{}
  &[\![71,12,d_e=15]\!]_2 & [\![36,12,8]\!]_2 & 14 \\\hline
8 &[\![49,1,d_e=17]\!]_2 & [\![25,1,9]\!]_2 & 13 \\{}
  &[\![53,3,d_e=17]\!]_2 & [\![27,3,9]\!]_2 & 13 \\{}
  &[\![69,4,d_e=17]\!]_2 & [\![35,4,9]\!]_2 & 15 \\{}
  &[\![101,19,d_e=17]\!]_2 & [\![51,19,9]\!]_2 & 16 \\\hline
9 &[\![55,2,d_e=19]\!]_2 & [\![28,2,10]\!]_2 & 14 \\{}
  &[\![71,3,d_e=19]\!]_2 & [\![36,3,10]\!]_2 & 15 \\{}
  &[\![105,17,d_e=19]\!]_2 & [\![53,17,10]\!]_2 & 17 \\\hline
10&[\![57,1,d_e=21]\!]_2 & [\![29,1,11]\!]_2 & 15 \\{}
  &[\![81,3,d_e=21]\!]_2 & [\![41,3,11]\!]_2 & 18 \\{}
  &[\![95,4,d_e=21]\!]_2 & [\![48,4,11]\!]_2 & 20 \\{}
  &[\![97,5,d_e=21]\!]_2 & [\![49,5,11]\!]_2 & 19 \\\hline
11 &[\![83,2,d_e=23]\!]_2 & [\![42,2,12]\!]_2 & 19 \\{}
  &[\![97,3,d_e=23]\!]_2 & [\![49,3,12]\!]_2 & 21 \\{}
  &[\![99,4,d_e=23]\!]_2 & [\![50,4,12]\!]_2 & 20 \\{}
  &[\![107,8,d_e=23]\!]_2 & [\![54,8,12]\!]_2 & 19 \\\hline
12 &[\![85,1,d_e=25]\!]_2 & [\![43,1,13]\!]_2 & 21 \\{}
  &[\![101,3,d_e=25]\!]_2 & [\![51,3,13]\!]_2 & 21 \\{}
  &[\![113,5,d_e=25]\!]_2 & [\![57,5,13]\!]_2 & 21 \\\hline
13 &[\![103,2,d_e=27]\!]_2 & [\![52,2,14]\!]_2 & 21 \\{}
  &[\![115,4,d_e=27]\!]_2 & [\![58,4,14]\!]_2 & 22 \\{}
  &[\![125,6,d_e=27]\!]_2 & [\![63,6,14]\!]_2 & 23 \\\hline
14 &[\![105,1,d_e=29]\!]_2 & [\![53,1,15]\!]_2 & 21 \\{}
  &[\![117,3,d_e=29]\!]_2 & [\![59,3,15]\!]_2 & 23 \\\hline
15 &[\![119,2,d_e=31]\!]_2 & [\![60,2,16]\!]_2 & 23 \\\hline
16 &[\![121,1,d_e=33]\!]_2 & [\![61,1,17]\!]_2 & 25 \\\hline
\end{array}$}
\end{table}

\begin{table}[hbt]
\caption{Concatenated codes $[\![n,k,d_e]\!]_2$ for the AD channel
  based on non-QMDS outer codes with qudit dimension $4$ and $8$.}
\label{tab:nonQMDS4_8}\def\arraystretch{1.15}
\centerline{$
\begin{array}{|c|c|c|c|}
\hline
t & \text{concatenated code} & \text{outer code} & d_{\text{lb}}\\
\hline
4 &[\![41,8,d_e=9]\!]_2 & [\![14,4,5]\!]_{2^2} & 8 \\{}
  &[\![44,10,d_e=9]\!]_2 & [\![15,5,5]\!]_{2^2} & 8 \\{}
  &[\![47,12,d_e=9]\!]_2 & [\![16,6,5]\!]_{2^2} & 8 \\\hline
5 &[\![50,10,d_e=11]\!]_2 & [\![17,5,6]\!]_{2^2} & 9 \\\hline
6 &[\![44,2,d_e=13]\!]_2 & [\![15,1,7]\!]_{2^2} & 12 \\{}
  &[\![56,6,d_e=13]\!]_2 & [\![19,3,7]\!]_{2^2} & 12 \\{}
  &[\![59,8,d_e=13]\!]_2 & [\![20,4,7]\!]_{2^2} & 12 \\{}
  &[\![74,14,d_e=13]\!]_2 & [\![25,7,7]\!]_{2^2} & 12 \\{}
  &[\![77,16,d_e=13]\!]_2 & [\![26,8,7]\!]_{2^2} & 12 \\{}
  &[\![80,18,d_e=13]\!]_2 & [\![27,9,7]\!]_{2^2} & 12 \\\hline
7 &[\![104,26,d_e=15]\!]_2 & [\![35,13,8]\!]_{2^2} & 14 \\{}
  &[\![107,28,d_e=15]\!]_2 & [\![36,14,8]\!]_{2^2} & 14 \\\hline
8 &[\![74,6,d_e=17]\!]_2 & [\![25,3,9]\!]_{2^2} & 15 \\{}
  &[\![92,14,d_e=17]\!]_2 & [\![31,7,9]\!]_{2^2} & 16 \\{}
  &[\![95,16,d_e=17]\!]_2 & [\![32,8,9]\!]_{2^2} & 16 \\{}
  &[\![110,22,d_e=17]\!]_2 & [\![37,11,9]\!]_{2^2} & 16 \\\hline
9 &[\![77,4,d_e=19]\!]_2 & [\![26,2,10]\!]_{2^2} & 16 \\{}
  &[\![98,10,d_e=19]\!]_2 & [\![33,5,10]\!]_{2^2} & 18 \\{}
  &[\![101,12,d_e=19]\!]_2 & [\![34,6,10]\!]_{2^2} & 17 \\{}
  &[\![104,14,d_e=19]\!]_2 & [\![35,7,10]\!]_{2^2} & 17 \\{}
  &[\![110,18,d_e=19]\!]_2 & [\![37,9,10]\!]_{2^2} & 17 \\{}
  &[\![113,20,d_e=19]\!]_2 & [\![38,10,10]\!]_{2^2} & 18 \\{}
  &[\![116,22,d_e=19]\!]_2 & [\![39,11,10]\!]_{2^2} & 18 \\\hline
10&[\![95,4,d_e=21]\!]_2 & [\![32,2,11]\!]_{2^2} & 20 \\{}
  &[\![98,6,d_e=21]\!]_2 & [\![33,3,11]\!]_{2^2} & 19 \\{}
  &[\![101,8,d_e=21]\!]_2 & [\![34,4,11]\!]_{2^2} & 19 \\{}
  &[\![104,10,d_e=21]\!]_2 & [\![35,5,11]\!]_{2^2} & 18 \\{}
  &[\![107,12,d_e=21]\!]_2 & [\![36,6,11]\!]_{2^2} & 18 \\{}
  &[\![116,14,d_e=21]\!]_2 & [\![39,7,11]\!]_{2^2} & 20 \\{}
  &[\![119,16,d_e=21]\!]_2 & [\![40,8,11]\!]_{2^2} & 20 \\{}
  &[\![122,18,d_e=21]\!]_2 & [\![41,9,11]\!]_{2^2} & 20 \\{}
  &[\![125,20,d_e=21]\!]_2 & [\![42,10,11]\!]_{2^2} & 20 \\{}
  &[\![128,22,d_e=21]\!]_2 & [\![43,11,11]\!]_{2^2} & 20 \\\hline
11 &[\![116,10,d_e=23]\!]_2 & [\![39,5,12]\!]_{2^2} & 21 \\{}
  &[\![119,12,d_e=23]\!]_2 & [\![40,6,12]\!]_{2^2} & 21 \\{}
  &[\![122,14,d_e=23]\!]_2 & [\![41,7,12]\!]_{2^2} & 21 \\{}
  &[\![125,16,d_e=23]\!]_2 & [\![42,8,12]\!]_{2^2} & 21 \\\hline
12&[\![116,6,d_e=25]\!]_2 & [\![39,3,13]\!]_{2^2} & 22 \\{}
  &[\![119,8,d_e=25]\!]_2 & [\![40,4,13]\!]_{2^2} & 22 \\{}
  &[\![128,10,d_e=25]\!]_2 & [\![43,5,13]\!]_{2^2} & 23 \\\hline
\multicolumn{4}{c@{}}{\rule{0pt}{10pt}}\\
\hline
t & \text{concatenated code} & \text{outer code} & d_{\text{lb}}\\
\hline
8 & [\![107,21,d_e=17]\!]_2 & [\![27,7,9]\!]_{2^3} & 16 \\\hline
\end{array}$}
\end{table}

\section{Discussion}

We can also use other asymmetric codes as inner codes to construct
concatenated codes based on Theorem~\ref{th:con}.  Using a similar
idea as in Theorem~\ref{th:main}, one may also encode the first qudit
of the outer $[\![n_2,k_2]\!]_{2^{k_2}}$ code into a trivial
$[\![k_2,k_2]\!]_2$ code. This leads to the following corollary.

\begin{corollary}
\label{cor:con}
Concatenating an $[\![n_2,k_2,\delta]\!]_{2^{k_1}}$ qudit outer code
$\mathcal{Q}_{o}$ with an inner asymmetric $[\![n_1,k_1]\!]_2$ code
$\mathcal{Q}_{i}$ with effective distance $d_e$ results in a code
$[\![n_1(n_2-1)+k_2,k_1k_2]\!]_2$ with effective distance at least
$d_e(\delta-1)+1$, as well as a concatenated code
$[\![n_1n_2,k_1k_2]\!]_2$ with effective distance at least
$d_e\delta$.
\end{corollary}
\begin{example}
Choose the inner code to be the asymmetric $[\![8,3,\{4,2\}]\!]_2$ CSS
code with $X$-distance $d_X=4$ and $Z$-distance $d_Z=2$, resulting in
effective distance $d_e=4$.  It can be constructed from the first
order Reed-Muller code and the repetition code.  Its stabilizer is
generated by
\[\arraycolsep0.5\arraycolsep
\begin{array}{cccccccc}
Z & Z & Z & Z & I & I & I & I\\
Z & Z & I & I & Z & Z & I & I\\
Z & I & Z & I & Z & I & Z & I\\
Z & Z & Z & Z & Z & Z & Z & Z\\
X & X & X & X & X & X & X & X
\end{array}
\]
Based on Theorem~\ref{th:con}, concatenating with a QMDS
$[\![10,2,5]\!]_{2^3}$ outer code results in a code $[\![80,6]\!]_2$
with effective distance $d_e=20$.  This code is better than the best
known stabilizer code $[\![80,6,16]\!]_2$.  Using
Corollary~\ref{cor:con}, we get a $[\![75,6]\!]_2$ code with effective
distance $d_e=17$, correcting $t=8$ AD errors.  This again improves
upon the best known stabilizer code $[\![75,6,15]\!]_2$. However, the
$t=8$ code with parameters $[\![74,6]\!]_2$ listed in
Table~\ref{tab:nonQMDS4_8} has better parameters. Note that for both
codes $[\![8,3,\{4,2\}]\!]_2$ and $[\![2,1,\{2,1\}]\!]_2$ (i.e., the code 
$\mathcal{Q}_2$ with the stabilizer generated by $ZZ$), the ratio
between the $X$- and $Z$-distance is $2$, resulting in an effective
distance of $4$ and $2$, respectively. However, the $[\![2,1,\{2,1\}]\!]_2$ code has
rate $1/2$ compared to rate $3/8$ for the $[\![8,3,\{4,2\}]\!]_2$ code, resulting in
codes with better parameters.  

Nonetheless, this example illustrates the flexibility of our method.
We can also use it for channels for which the asymmetry between
$p_{xy}$ and $p_z$ is different than for the amplitude damping
channel (see, e.g.~\cite{PhysRevA.75.032345}).
\end{example}

\section*{Acknowledgements} TJ and BZ are supported by NSERC.



%

\bibliographystyle{IEEEtranS}
\bibliography{AD}

\begin{thebibliography}{10}
\providecommand{\url}[1]{#1}
\csname url@samestyle\endcsname
\providecommand{\newblock}{\relax}
\providecommand{\bibinfo}[2]{#2}
\providecommand{\BIBentrySTDinterwordspacing}{\spaceskip=0pt\relax}
\providecommand{\BIBentryALTinterwordstretchfactor}{4}
\providecommand{\BIBentryALTinterwordspacing}{\spaceskip=\fontdimen2\font plus
\BIBentryALTinterwordstretchfactor\fontdimen3\font minus
  \fontdimen4\font\relax}
\providecommand{\BIBforeignlanguage}[2]{{%
\expandafter\ifx\csname l@#1\endcsname\relax
\typeout{** WARNING: IEEEtranS.bst: No hyphenation pattern has been}%
\typeout{** loaded for the language `#1'. Using the pattern for}%
\typeout{** the default language instead.}%
\else
\language=\csname l@#1\endcsname
\fi
#2}}
\providecommand{\BIBdecl}{\relax}
\BIBdecl

\bibitem{bennett1996mixed}
C.~H. Bennett, D.~P. DiVincenzo, J.~A. Smolin, and W.~K. Wootters,
  ``Mixed-state entanglement and quantum error correction,'' \emph{Physical
  Review A}, vol.~54, no.~5, pp. 3824--3851, 1996.

\bibitem{chuang1997bosonic}
I.~L. Chuang, D.~W. Leung, and Y.~Yamamoto, ``Bosonic quantum codes for
  amplitude damping,'' \emph{Physical Review A}, vol.~56, no.~2, pp.
  1114--1125, 1997.

\bibitem{CSS+09}
A.~Cross, G.~Smith, J.~A. Smolin, and B.~Zeng, ``Codeword stabilized quantum
  codes,'' \emph{IEEE Transactions on Information Theory}, vol.~55, no.~1, pp.
  433--438, 2009.

\bibitem{duan2010multi}
R.~Duan, M.~Grassl, Z.~Ji, and B.~Zeng, ``Multi-error-correcting amplitude
  damping codes,'' in \emph{Proceedings 2010 IEEE International Symposium on
  Information Theory (ISIT 2010)}, 2010, pp. 2672--2676.

\bibitem{fletcher2008channel}
A.~S. Fletcher, P.~W. Shor, and M.~Z. Win, ``Channel-adapted quantum error
  correction for the amplitude damping channel,'' \emph{IEEE Transactions on
  Information Theory}, vol.~54, no.~12, pp. 5705--5718, 2008.

\bibitem{thesis:gottesman}
D.~Gottesman, ``Stabilizer codes and quantum error correction,'' Ph.D.
  dissertation, California Institute of Technology, Pasadena, CA, 1997.

\bibitem{Grassl:codetables}
M.~Grassl, ``{Bounds on the minimum distance of linear codes and quantum
  codes},'' Online available at \url{http://www.codetables.de}, 2007, accessed
  on 2016-01-18.

\bibitem{GBR04}
M.~Grassl, T.~Beth, and M.~R{\"o}tteler, ``On optimal quantum codes,''
  \emph{International Journal of Quantum Information}, vol.~2, no.~1, pp.
  55--64, 2004.

\bibitem{GrRo15}
M.~Grassl and M.~R{\"o}tteler, ``Quantum {MDS} codes over small fields,'' in
  \emph{Proceedings 2015 IEEE International Symposium on Information Theory
  (ISIT 2015)}, 2015, pp. 1104--1108.

\bibitem{GWYZ14}
M.~Grassl, Z.~Wei, Z.-Q. Yin, and B.~Zeng, ``Quantum error-correcting codes for
  amplitude damping,'' in \emph{Proceedings 2014 IEEE International Symposium
  on Information Theory (ISIT 2014)}, 2014, pp. 906--910.

\bibitem{PhysRevA.75.032345}
L.~Ioffe and M.~M{\'e}zard, ``Asymmetric quantum error-correcting codes,''
  \emph{Physical Review A}, vol.~75, no.~3, p. 032345, Mar 2007.

\bibitem{JGZ16}
T.~Jackson, M.~Grassl, and B.~Zeng, ``Codeword stabilized quantum codes for
  asymmetric channels,'' 2016, arXiv:1601.05763 [quant-ph].

\bibitem{klove1981error}
T.~Kl{\o}ve, \emph{Error correcting codes for the asymmetric channel}.\hskip
  1em plus 0.5em minus 0.4em\relax Department of Pure Mathematics, University
  of Bergen, 1981.

\bibitem{knill1997theory}
E.~Knill and R.~Laflamme, ``Theory of quantum error-correcting codes,''
  \emph{Physical Review A}, vol.~55, no.~2, pp. 900--911, 1997.

\bibitem{KLZ98}
E.~Knill, R.~Laflamme, and W.~H. Zurek, ``{Resilient quantum computation: error
  models and thresholds},'' \emph{Proceedings of the Royal Society of London
  Series~A}, vol. 454, no. 1969, pp. 365--384, 1998.

\bibitem{laflamme1996perfect}
R.~Laflamme, C.~Miquel, J.~P. Paz, and W.~H. Zurek, ``Perfect quantum error
  correcting code,'' \emph{Physical Review Letters}, vol.~77, no.~1, p. 198,
  1996.

\bibitem{lang2007nonadditive}
R.~Lang and P.~W. Shor, ``Nonadditive quantum error correcting codes adapted to
  the amplitude damping channel,'' 2007, arXiv:0712.2586 [quant-ph].

\bibitem{leung1997approximate}
D.~W. Leung, M.~A. Nielsen, I.~L. Chuang, and Y.~Yamamoto, ``Approximate
  quantum error correction can lead to better codes,'' \emph{Physical Review
  A}, vol.~56, no.~4, pp. 2567--2573, 1997.

\bibitem{nielsenchuang}
M.~Nielsen and I.~Chuang, \emph{Quantum computation and quantum
  information}.\hskip 1em plus 0.5em minus 0.4em\relax Cambridge, England:
  Cambridge University Press, 2000.

\bibitem{Rai99}
E.~M. Rains, ``Nonbinary quantum codes,'' \emph{IEEE Transactions on
  Information Theory}, vol.~45, no.~6, pp. 1827--1832, 1999.

\bibitem{shor2011high}
P.~W. Shor, G.~Smith, J.~Smolin, and B.~Zeng, ``High performance
  single-error-correcting quantum codes for amplitude damping,'' \emph{IEEE
  Transactions on Information Theory}, vol.~57, no.~10, pp. 7180--7188, 2011.

\end{thebibliography}

\end{document}